\documentclass[prl,twocolumn,showpacs,superscriptaddress]{revtex4}
\usepackage[dvips]{epsfig}
\usepackage{graphicx}
\usepackage{amsfonts}
\usepackage{bm}
\usepackage{amsmath,amssymb,lscape,float}

\makeatletter
\renewcommand\appendix{\par
  \setcounter{section}{0}%
  \setcounter{subsection}{0}%
  \setcounter{equation}{0}%
  \gdef\thesection{\Alph{section}}
  \@addtoreset {equation}{section}
  \renewcommand{\theequation}{\thesection\arabic{equation}}}
\makeatother

\begin{document}
\title{Electron-phonon coupling in crystalline organic semiconductors:\\
       Microscopic evidence for nonpolaronic charge carriers}
\author{Nenad Vukmirovi\'c}
\email{nenad.vukmirovic@ipb.ac.rs}
\affiliation{Scientific Computing Laboratory, Institute of Physics Belgrade, University of Belgrade,
             Pregrevica 118, 11080 Belgrade, Serbia}
\author{C. Bruder}
\affiliation{Department of Physics, University of Basel,
Klingelbergstrasse 82, CH-4056 Basel, Switzerland}
\author{Vladimir M. Stojanovi\'c}
\email{vladimir.stojanovic@unibas.ch}
\affiliation{Department of Physics, University of Basel,
 Klingelbergstrasse 82, CH-4056 Basel, Switzerland}

\date{\today}
\begin{abstract}
We consider electron(hole)-phonon coupling in crystalline organic semiconductors, using naphthalene 
for our case study. Employing a first-principles approach, we compute the changes in the self-consistent 
Kohn-Sham potential corresponding to different phonon modes and go on to obtain the  
carrier-phonon coupling matrix elements (vertex functions). We then evaluate perturbatively the quasiparticle 
spectral residues for electrons at the bottom of the lowest-unoccupied- (LUMO) and holes at the top of the 
highest-occupied (HOMO) band, respectively obtaining $Z_{\textrm{e}}\approx 0.74$ and $Z_{\textrm{h}}\approx 0.78$. 
Along with the widely accepted notion that the carrier-phonon coupling strengths in polyacenes 
decrease with increasing molecular size, our results provide a strong microscopic evidence 
for the previously conjectured nonpolaronic nature of band-like carriers in these systems.
\end{abstract}
\maketitle

High-purity crystalline organic semiconductors, best exemplified by polyacenes, have attracted 
a great deal of interest as they hold promise for plastic electronics~\cite{KarlIn2001,Giri++:11}. Despite many 
theoretical~\cite{OrganicCrystalsTheory,Kenkre+:89,Hanewald++:04,HanewaldEtAlJPhys,Cheng+Silbey:08,Picon+Stehr+Perroni} 
and experimental~\cite{Warta+Karl:85,Gershenson++:06,OrgSemExperiment} studies, a 
controversy persists regarding the nature of charge carriers in these $\pi$-electron 
systems with weak van der Waals intermolecular interactions. Are these carriers 
of polaronic character or not? Elucidating this important issue would pave the way for understanding 
the intrinsic (bulk) charge transport in polyacenes and transport in the organic field-effect-transistor 
geometry~\cite{Gershenson++:06}. While recent theoretical studies have provided a qualitative 
case for non-polaronic carriers~\cite{TroisiEtAl,FratiniAndCiuchi}, 
this conclusion has yet to be borne out through a detailed microscopic
analysis of the underlying electron(hole)-phonon (e-ph) coupling.

While the e-ph coupling in inorganic semiconductors is rather weak, in their 
narrow-band organic counterparts (the conduction- and valence bandwidths in polyacenes are 
$W\sim 0.1-0.4$\:eV) it plays a prominent role~\cite{Coropceanu++:02,Girlando++:10}. 
Like in other $\pi$-electron systems, this interaction has a strong momentum 
dependence stemming from Peierls' coupling mechanism~\cite{Zoli+Stojanovic,Li++:12}, whereby lattice 
displacements affect electronic hopping integrals. Given their structural complexity and pronounced 
anisotropy, e-ph coupling in organic semiconductors is typically described using tight-binding-type 
models, for simplicity often taken to be one-dimensional and merely involving Einstein phonons. 
Although such simplified models often lead to qualitatively correct results, they cannot be considered 
quantitatively reliable as even the bare band structure of polyacenes can only be accurately reproduced
if at least three different hopping integrals are taken into account~\cite{Yoshida+Sato:08}.

\begin{figure}[b!]
\includegraphics[width=7.5cm]{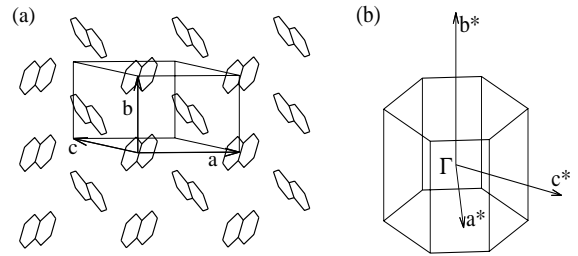}
\caption{\label{NaphthIllustr} (a) An illustration of the crystal structure of naphthalene, 
comprising molecular layers (in the $a-b$ plane) stacked in the crystallographic $c$ direction. 
Each unit cell contains two inequivalent naphthalene molecules in a herringbone-like arrangement. 
(b) The irreducible wedge of the Brillouin zone in the naphthalene crystal. The zone center 
($\bm{\Gamma}$-point) is indicated, along with the reciprocal basis vectors.}
\end{figure}

In this Letter, on the example of the naphthalene crystal [see Fig.~\ref{NaphthIllustr}(a)], 
we systematically investigate e-ph coupling in organic semiconductors. As a by-product of 
an electronic-structure calculation performed within the density functional theory (DFT) 
framework, we compute the changes in the self-consistent Kohn-Sham potential 
for electrons in the LUMO- and holes in the HOMO band coupled to different optical phonon modes. 
By sampling the e-ph scattering processes throughout the Brillouin zone (BZ) 
[Fig.~\ref{NaphthIllustr}(b)], we extract the corresponding (momentum-dependent) coupling 
matrix elements~\cite{Giustino++} and make use of the latter to perturbatively compute the quasiparticle 
spectral residues and inelastic scattering rates.
In particular, the values we obtain for the quasiparticle residues of electrons at the bottom of the LUMO band 
($Z_{\textrm{e}}\approx 0.74$) and holes at the top of the HOMO band ($Z_{\textrm{h}}\approx 0.78$) 
indicate that the band-like carriers in naphthalene do not have polaronic character. 
This conclusion can be carried over to higher polyacenes (anthracene, tetracene, pentacene) in which
e-ph coupling becomes progressively weaker. 

We make use of the \texttt{Quantum-ESPRESSO}~\cite{QuantumEspresso} code, with
the core-valence interaction taken into account through a norm-conserving
pseudopotential with kinetic energy cutoff of $60$ Ry. Our first-principles scheme 
employs the PBE GGA exchange-correlation functional with semiempirical correction 
for van der Waals interaction~\cite{GrimmeJCC:06}. The functionals specifically 
tailored for this type of weak intermolecular interaction have proven to yield 
accurate cohesive energies of polyacenes~\cite{AmbroschDraxl++}. Our calculations 
of the phonon spectrum and e-ph matrix elements are based on density-functional 
perturbation theory~\cite{Baroni++:01}.

The momentum-space form of the most general (multi-band) e-ph-coupling 
Hamiltonian reads
\begin{equation}\label{mscoupling}
\hat{H}_{\textrm{e-ph}}=\frac{1}{\sqrt{N}}
\sum_{nn',\mathbf{k,q},\lambda}\gamma_{n'n}^{\lambda}(\mathbf{k,q}) \:
\hat{a}_{n',\mathbf{k+q}}^{\dagger}\hat{a}_{n,\mathbf{k}}
(\hat{b}_{-\mathbf{q},\lambda}^{\dagger}+\hat{b}_{\mathbf{q},\lambda}) \:,
\end{equation}
\noindent where $\hat{a}_{n,\mathbf{k}}$ destroys an electron with quasimomentum 
$\mathbf{k}$ in the $n$-th Bloch band, $\hat{b}_{\mathbf{q},\lambda}$ 
a phonon of branch $\lambda$ with quasimomentum $\mathbf{q}$ 
(frequency $\omega_{\lambda,\mathbf{q}}$), and 
\begin{eqnarray}\label{vertfunc}
\gamma_{n'n}^{\lambda}(\mathbf{k,q})&=&\sqrt{\frac{\hbar}{2\omega_{\lambda,\mathbf{q}}}}
\sum_{S\alpha}e_{S\alpha}^{(\lambda)}(\mathbf{q})\frac{1}{\sqrt{M_S}}\times \nonumber\\ && \times
\left\langle\psi_{n',\mathbf{k+q}}  \left|
\frac{\partial U_{\textrm{scf}}}{\partial u_{\mathbf{q}S\alpha}}\right|\:\psi_{n,\mathbf{k}}\right
\rangle
\end{eqnarray}
\noindent stand for the (bare) e-ph interaction vertex functions.
In the last expression, $|\psi_{n\mathbf{k}}\rangle$ and $|\psi_{n',\mathbf{k+q}}\rangle$
are electronic Bloch states, $U_{\textrm{scf}}$ is
the self-consistent Kohn-Sham potential, $e_{S\alpha}^{(\lambda)}(\mathbf{q})$ is the component
of the eigenvector of the dynamical matrix $D(\mathbf{q})$ corresponding to the displacement of the $S$-th
atom (mass $M_{S}$) in direction $\alpha$ due to the phonon mode ($\lambda,\mathbf{q}$),
while $u_{\mathbf{q}S\alpha}\equiv N^{-1}\sum_p \exp{\left(i{\mathbf{q}}\cdot \mathbf{R}_p\right)}u_{pS\alpha}$,
where $\mathbf{R}_{p}$ ($p=1,\ldots,N$) denotes unit cells and $u_{pS\alpha}$ the displacement 
of the atom $S$ from unit cell $p$ in direction $\alpha$.

The crux of our treatment is a first-principles evaluation of the vertex functions
in Eq.~\eqref{vertfunc} for two pairs of HOMO- and LUMO-derived bands, respectively 
denoted by (HOMO$-1$, HOMO), and (LUMO, LUMO$+1$). These functions incorporate the 
contributions of the corresponding phonon modes to all the existing e-ph coupling 
mechanisms. The actual momentum dependence that characterizes the coupling to a given phonon 
mode depends in a complex way on the underlying molecular-crystal geometry, the symmetry of 
the mode involved, as well as the spatial directionality of the $\pi$-electron 
orbitals~\cite{Stojanovic+:10}. Such aspects clearly cannot be accounted for within 
simplified (low-dimensional) tight-binding-type models, hence providing a rationale 
for our rigorous -- albeit computationally much more demanding -- approach.

The electronic structure and phonon spectra calculations were
performed on a uniform $6\times 6\times 6$ momentum grid. The changes of
$U_{\textrm{scf}}$ with respect to atomic-displacement perturbations
were obtained using the \texttt{Phonon} code from the
\texttt{Quantum-ESPRESSO}~\cite{QuantumEspresso} package, while the
matrix elements in Eq.~\eqref{vertfunc} were extracted from the
\texttt{EPW} code~\cite{Noffsinger++}. Our own implementation of the
Fourier-Wannier interpolation scheme for e-ph matrix
elements~\cite{Giustino++} was then used to obtain the matrix elements
on a dense $24\times 24\times 24$ grid, which is necessary
for achieving high accuracy when computing $\mathbf{q}$-space
integrals. The \texttt{Wannier90} code~\cite{SouzaMostofi} was employed to get
the maximally-localized Wannier functions needed for this
interpolation scheme. While this methodology has heretofore been used
for systems with few atoms per unit cell~\cite{Giustino++}, our
implementation enabled its application to a system with as many as
$36$ atoms.
\begin{figure}[b!]
\includegraphics[width=6cm]{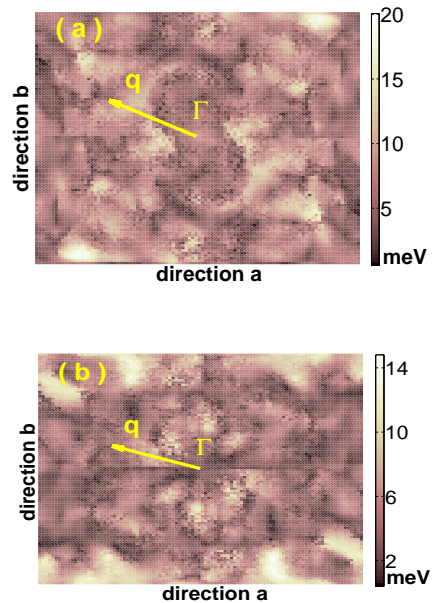}
\caption{\label{VFinter}Calculated phonon-momentum dependence of the moduli of the e-ph 
vertex functions in the $a-b$ plane for (a) electrons at the bottom of the LUMO band, and 
(b) holes at the top of the HOMO band, interacting with phonons of the lowest optical branch
(the zone-center energy is $11.4$ meV).}
\end{figure}

We computed the e-ph vertex functions for different phonon branches $\lambda$ 
and quasimomenta $\mathbf{k}$, $\mathbf{q}$ throughout the BZ. The strong momentum dependence and 
the anisotropic character of these functions, especially in the $a-b$ plane, is illustrated in 
Fig.~\ref{VFinter} for electrons at the bottom of the LUMO band ($\mathbf{k}=0$) and holes at the 
top of the HOMO band ($\mathbf{k}=0.5\:\mathbf{a}^{*}+0.5\:\mathbf{b}^{*}$) interacting with the lowest-branch 
(intermolecular) optical phonons of rotational (librational) origin. This strong momentum dependence 
serves as evidence for the dominance of the Peierls-type (off-diagonal) e-ph coupling over 
the strictly local Holstein-type coupling, as the latter yields completely momentum-independent 
vertex functions. 
\begin{figure}[b!]
\includegraphics[width=7.6cm]{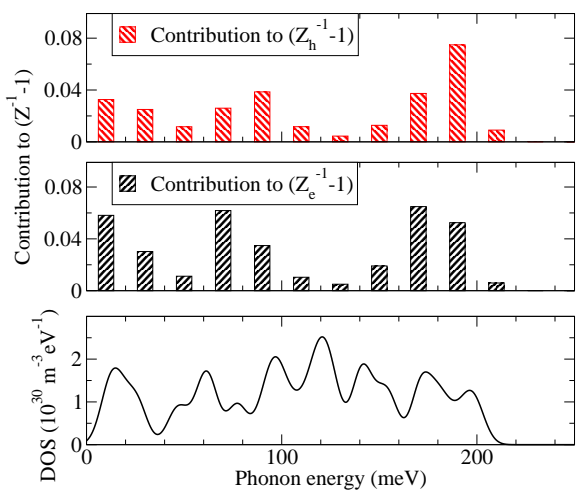}
\caption{\label{histograms}Upper panel: Contributions of different phonon modes, divided 
into $20$\:meV-wide energy intervals, to the quasiparticle renormalization for 
electrons and holes. Lower panel: Phonon density-of-states (DOS).}
\end{figure}
Besides, contrary to what is often assumed in the literature, our calculations show 
that the coupling to the high-energy intramolecular phonon modes is not necessarily 
much weaker than the coupling to the intermolecular ones.  
Among the intramolecular branches the most relevant ones are two pairs with 
the respective zone-center energies of approximately $172$\:meV and $194$\:meV.

The renormalization due to e-ph coupling is characterized by the quasiparticle 
spectral residue $Z_{n}(\mathbf{k})\equiv|\langle\Psi_{n\mathbf{k}}|\psi_{n\mathbf{k}}\rangle|^{2}$\:,
where $|\psi_{n\mathbf{k}}\rangle$ is a bare-electron (or hole) state in the $n$-th Bloch band 
and $|\Psi_{n\mathbf{k}}\rangle$ that of the coupled e-ph system. Our calculations show that 
the average values of the momentum-dependent vertex functions (over all the phonon branches and 
quasimomenta) are approximately $6.8$\:meV for the HOMO-band holes and $7.8$\:meV for the LUMO-band 
electrons. These average values, i.e., their relative smallness compared to the HOMO- and 
LUMO bandwidths ($234.3$ and $207.6$\:meV, respectively), justify a perturbative evaluation 
of the quasiparticle residues. Within the framework of Rayleigh-Schr\"{o}dinger (RS) perturbation 
theory, in the lowest nonvanishing (second) order one obtains
\begin{equation}\label{Zcinverse}
Z_{n}^{-1}(\mathbf{k})=1+\frac{1}{N}\sum_{n',\mathbf{q},\lambda}
\frac{|\gamma_{n'n}^{\lambda}(\mathbf{k},\mathbf{q})|^{2}}{\big[\varepsilon_{n}
(\mathbf{k})-\varepsilon_{n'}(\mathbf{k}+\mathbf{q})
-\hbar\omega_{\lambda,\mathbf{q}}\big]^{2}} \:,
\end{equation}
where $\varepsilon_{n}(\mathbf{k})$ is the band dispersion.

Given the pairs of energetically close HOMO- and LUMO-derived bands, 
for electrons at the bottom of the LUMO band the interband 
contributions [the terms with $n'\neq n$ in Eq.~\eqref{Zcinverse}] originate 
from the LUMO+1 band, while for holes at the top of the HOMO band 
such contributions stem from its HOMO-1 counterpart. We obtain $Z_{\textrm{e}}\approx 0.74$ 
and $Z_{\textrm{h}}\approx 0.78$ for the quasiparticle residues of electrons and holes, respectively.
The contributions of different phonon modes to the quasiparticle
renormalization are shown in Fig.~\ref{histograms}, which illustrates
the prominent role of high-energy intramolecular phonons for both
electrons and holes. From RS perturbation theory, we also 
find the carrier binding (relaxation) energies: $E_{\textrm{b}}=68.7$\:meV ($58.8$\:meV) 
for electrons (holes).

To check whether the e-ph coupling in the system is weak enough for
second-order RS perturbation theory to be reliable, we also computed $Z_{\textrm{e}}$ 
using the self-consistent Born approximation (SCBA). The SCBA self-energy is based on 
all second-order diagrams and an infinite (non-crossing) subset of higher-order ones~\cite{Stojanovic+:10}. 
Given that the SCBA approach is computationally extremely demanding, the calculation was done 
using the original $6\times 6\times 6$ grid and taking only the LUMO band into account. 
The obtained result $Z_{\textrm{e}}\approx 0.78$ agrees rather well with the RS result 
obtained on the same grid and with only intraband processes taken into account 
($Z_{\textrm{e}}\approx 0.77$, slightly larger than $Z_{\textrm{e}}\approx 0.74$ 
obtained on a dense $24\times 24\times 24$ grid). The agreement between the two 
values suggests that we are in the weak-coupling regime where RS perturbation 
theory is as accurate as more sophisticated approaches.

The fact that $Z_{\textrm{h}} > Z_{\textrm{e}}$ indicates that the carrier-phonon coupling is 
somewhat stronger for electrons than for holes, thus corroborating the conclusions of some 
earlier studies~\cite{Coropceanu++:02}. Furthermore, the rather large values obtained for
$Z_{\textrm{e}}$ and $Z_{\textrm{h}}$ suggest that the phonon-induced renormalization is insufficient 
for these carriers to have polaronic character. While a direct comparison to polaron models with dispersionless 
phonons is not possible, let us mention that, e.g., in the Holstein model $Z=0.1-0.2$ at the onset of
self trapping (the actual value of $Z$ depends on the dimensionality and the ratio of the
hopping integral and the relevant phonon energy)~\cite{ZoliReview}. Another argument in favor of non-polaronic 
carriers comes from the comparison of their binding energies and the half-bandwiths: the criterion for 
small (lattice) polaron~\cite{AlexandrovDevreese} formation $E_{\textrm{b}}\geq W/2$ is neither
satisfied for electrons nor for holes. While our present analysis is restricted to naphthalene, 
the last conclusion about the non-polaronic nature of carriers can be generalized to higher polyacenes.
The e-ph coupling strengths in these systems are known to decrease (while the HOMO- and LUMO bandwidths 
increase) with increasing molecular size, thus disfavoring polaron formation. 

Quite generally, the total inelastic scattering rate (inverse scattering time) for an electron 
(or a hole) with quasimomentum $\mathbf{k}$ in the $n$-th Bloch band is given by
\begin{multline}\label{scatt_rate}
\left(\frac{1}{\tau}\right)_{n\mathbf{k}}=\frac{2\pi}{N\hbar}\sum_{n',\mathbf{q},\lambda}
|\gamma_{n'n}^{\lambda}(\mathbf{k,q})|^{2}\left(\Delta^{\lambda,-}_{n'n}
+\Delta^{\lambda,+}_{n'n}\right)\:,
\end{multline}
where both the intraband ($n'=n$) and interband ($n'\neq n$) carrier
scattering processes into states with quasimomenta $\mathbf{k}+\mathbf{q}$ are taken into 
account, and
\begin{equation}
\Delta^{\lambda,\pm}_{n'n}\equiv\:(n_{\lambda,\mathbf{q}}+1/2\:\pm 1/2)\delta
(\varepsilon_{n',\mathbf{k+q}}-\varepsilon_{n,\mathbf{k}}\pm\hbar\omega_{\lambda,\mathbf{q}})
\end{equation}
corresponds to the emission ($+$) or absorption ($-$) of a phonon $(\lambda,\mathbf{q})$;
$n_{\lambda,\mathbf{q}}\equiv[\exp(\hbar\omega_{\lambda,\mathbf{q}}/k_{B}T)-1 ]^{-1}$ 
are the phonon occupation numbers at temperature $T$.

\begin{figure}[t!]
\includegraphics[width=6.5cm]{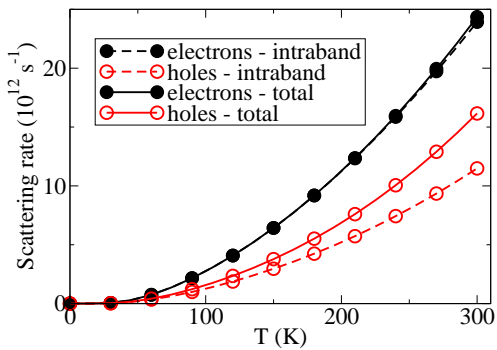}
\caption{\label{scatt_figure}The temperature dependence of  
inelastic scattering rates for band-like electron- and hole states in naphthalene.}
\end{figure}

Based on Eq.~\eqref{scatt_rate} and the calculated vertex functions, we evaluate the scattering 
rates at different temperatures for an electron at the bottom of the LUMO band and a hole at the top of the HOMO band, with 
the interband contributions originating from LUMO$+1$ and HOMO$-1$ bands, respectively.
The results, displayed in Fig.~\ref{scatt_figure}, show that for electrons the interband-scattering 
contribution is practically negligible, while for holes it is rather significant. This can be understood 
by noting that the HOMO and HOMO-1 bands have a large overlap through much of the BZ~\cite{Yoshida+Sato:08}. 
Importantly, from Fig.~\ref{scatt_figure} it can be inferred that even at  
room temperature the values of $\hbar/\tau$ for both electrons and holes 
are sufficiently smaller than the corresponding bandwidths that we can talk 
about well-defined, weakly renormalized, Bloch states. 

Let us now put our results in perspective by clarifying their connection to the
existing theories of crystalline organic semiconductors.
The long-held consensus that the conventional Bloch-Boltzmann theory
is inadequate for describing transport properties of these systems~\cite{OrganicCrystalsTheory}, 
as well as the awareness of the relevance of e-ph interaction, have led workers 
in the field to adopt polaron models~\cite{Kenkre+:89,Hanewald++:04,HanewaldEtAlJPhys,Cheng+Silbey:08}. 
The appeal to such models was partly motivated by the apparent mean free paths of carriers in these
systems being even below the intermolecular distance (the Mott-Ioffe-Regel limit). 
Yet, Troisi {\em et al.}~\cite{TroisiEtAl} put forward the idea of a temperature-driven dynamic 
disorder of these soft materials, which calls into question the applicability of the band concept 
but nonetheless allows one to reproduce the metallic-like power-law temperature dependence of the 
charge mobility seen in experiments. The characteristic duality between the band-like and incoherent carrier states 
was finally explained by Fratini and Ciuchi in their semi-phenomenological model~\cite{FratiniAndCiuchi}. 
One of the crucial assumptions underpinning this model is that the band-like states are only weakly
renormalized by the interaction with lattice vibrations. This weak renormalization, indicating
the absence of polarons, is quantitatively demonstrated by our present microscopic study. 

To conclude, on the example of naphthalene we have studied carrier-phonon coupling
in crystalline organic semiconductors. By combining a rigorous first-principles evaluation 
of coupling matrix elements and a perturbative many-body analysis, we established
that the band-like carrier states in these systems do not undergo a sufficiently strong 
phonon-induced renormalization to have polaronic character. Our study thus provides
a useful, up to now unavailable, microscopic insight into the nature of charge carriers 
in organic molecular crystals. 

\begin{acknowledgments}
Useful discussions with F. Giustino are gratefully aknowledged. 
N.V. was supported by European Community FP7 Marie Curie Career Integration Grant (ELECTROMAT), 
the Serbian Ministry of Science (project ON171017) and FP7 projects PRACE-1IP, PRACE-2IP, HP-SEE and EGI-InSPIRE.
V.M.S. and C.B. were supported by the Swiss NSF and the NCCR Nanoscience.
\end{acknowledgments}

\end{document}